\documentclass[aip,jap,reprint]{revtex4-1}
\usepackage[T1]{fontenc}
\usepackage{times}
\usepackage{graphicx,color}
\usepackage{amsfonts,amsmath,amssymb,amsbsy}
\usepackage[colorlinks=true, linkcolor=blue, citecolor=blue, urlcolor=blue]{hyperref}
\let\rm=\text
\let\bi=\boldsymbol
\let\mathbf=\boldsymbol
\def\emph#1{\textcolor{red}{#1}}
\def\emph#1{\textcolor{black}{#1}}

\begin{document}

\title{\emph{A microwave field-driven transistor-like skyrmionic device with the microwave current-assisted skyrmion creation}}

\author{Jing Xia}
\affiliation{School of Science and Engineering, The Chinese University of Hong Kong, Shenzhen 518172, China}

\author{Yangqi Huang}
\affiliation{Fert Beijing Institute, BDBC, and School of Electronic and Information Engineering, Beihang University, Beijing 100191, China}

\author{Xichao Zhang}
\affiliation{School of Science and Engineering, The Chinese University of Hong Kong, Shenzhen 518172, China}

\author{Wang Kang}
\affiliation{Fert Beijing Institute, BDBC, and School of Electronic and Information Engineering, Beihang University, Beijing 100191, China}

\author{Chentian Zheng}
\affiliation{Fert Beijing Institute, BDBC, and School of Electronic and Information Engineering, Beihang University, Beijing 100191, China}

\author{Xiaoxi Liu}
\affiliation{Department of Electrical and Computer Engineering, Shinshu University, 4-17-1 Wakasato, Nagano 380-8553, Japan}

\author{\\ Weisheng Zhao}
\email[E-mail:~]{weisheng.zhao@buaa.edu.cn}
\affiliation{Fert Beijing Institute, BDBC, and School of Electronic and Information Engineering, Beihang University, Beijing 100191, China}

\author{Yan Zhou}
\email[E-mail:~]{zhouyan@cuhk.edu.cn}
\affiliation{School of Science and Engineering, The Chinese University of Hong Kong, Shenzhen 518172, China}

\date{3 October 2017}

\begin{abstract}
\emph{Magnetic skyrmion is a topologically protected domain-wall structure at nanoscale, which could serve as a basic building block for advanced spintronic devices. Here, we propose a microwave field-driven skyrmionic device with the transistor-like function, where the motion of a skyrmion in a voltage-gated ferromagnetic nanotrack is studied by micromagnetic simulations. It is demonstrated that the microwave field can drive the motion of a skyrmion by exciting propagating spin waves, and the skyrmion motion can be governed by a gate voltage. We also investigate the microwave current-assisted creation of a skyrmion to facilitate the operation of the transistor-like skyrmionic device on the source terminal. It is found that the microwave current with an appropriate frequency can reduce the threshold current density required for the creation of a skyrmion from the ferromagnetic background. The proposed transistor-like skyrmionic device operated with the microwave field and current could be useful for building future skyrmion-based circuits.}
\end{abstract}


\maketitle

\section{Introduction}
\label{se:Introduction}
The magnetic skyrmion is an exotic magnetic texture which has a nanoscale vortex-like magnetization structure that is protected by topological invariance~\cite{Roszler_NATURE2006,Nagaosa_NNANO2013,Bergmann_JPCM2014,Wanjun_PHYSREP2017}. In the recent years, the magnetic skyrmion has been experimentally observed in magnetic materials~\cite{Muhlbauer_SCIENCE2009,Yu_NATURE2010,Heinze_NPHYS2011,Yu_NMATER2011,Schulz_NPHYS2012,Rimming_SCIENCE2013,Oike_NPHYS2015,Du_NCOMMS2015,Nii_NCOMMS2015,Buttner_NPHYS2015,Wanjun_SCIENCE2015,Wanjun_NATPHYS2017,Woo_ARXIV2015}, semiconductors~\cite{Kezsmarki_NMATER2015}, multiferroic~\cite{Seki_SCIENCE2012} and ferroelectric~\cite{Nahas_NCOMMS2015} materials, which indicates that there are diverse potential applications of magnetic skyrmions in the field of spintronics. Indeed, the magnetic skyrmion as information carrier attracts increasing interest for developing the next-generation data storage devices due to its remarkable stability, extremely small size, and low-current depinning property~\cite{Fert_NNANO2013,Sampaio_NNANO2013,Iwasaki_NNANO2013,Tomasello_SREP2014,Yan_NCOMMS2014,Xichao_NCOMMS2015,Xichao_SREP2015A,Koshibae_JJAP2015,Dai_SREP2014,Beg_SREP2015}. Concurrently, many other applications of the magnetic skyrmions have been proposed and demonstrated, such as skyrmion-based logic devices~\cite{Xichao_SREP2015B}, oscillators~\cite{Yan_NCOMMS2015,Senfu_NJP2015}, and electronic devices~\cite{Xichao_SREP2015C,Upadhyaya_PRB2015,PIEEE2016}.

For the practical applications of the magnetic skyrmion, it is required to generate and control the isolated magnetic skyrmion in a feasible way. The magnetic skyrmion can be generated with some kinds of energy injection, such as by applying a spin-polarized current~\cite{Sampaio_NNANO2013,Wanjun_SCIENCE2015,Wanjun_NATPHYS2017}, a local heating~\cite{Koshibae_NCOMMS2014}, and a laser~\cite{Finazzi_PRL2013}. The magnetic skyrmion can also be nucleated with the nano-patterning~\cite{Sun_PRL2013} and be converted from a domain-wall pair~\cite{Yan_NCOMMS2014}. For the manipulation of the magnetic skyrmion, one can use the spin current~\cite{Fert_NNANO2013,Iwasaki_NNANO2013}, the spin wave~\cite{Iwasaki_PRB2014,Schutte_PRB2014,Xichao_NANOTECH2015,Fusheng_NANOLETT2015}, as well as the thermal gradient~\cite{Everschor_PRB2012,Kong_PRL2013,Lin_PRL2014} to drive the magnetic skyrmion.

Recently, the skyrmion-based transistor-like device has been proposed and studied in Refs.~\onlinecite{Xichao_SREP2015C,Upadhyaya_PRB2015}, in which the magnetic skyrmion is driven by spin current in a voltage-gated nanotrack. The magnetic vortex also has been used to mimic the transistor-like device.~\cite{Barman_SCIREP2014} In this paper, we study the magnetic skyrmion transistor-like device operated with microwaves, that is, the microwave-field-driven motion and microwave-current-assisted nucleation of the magnetic skyrmion in a voltage-gated nanotrack, where the perpendicular magnetic anisotropy (PMA) of the gate region is controlled by the applied voltage. The microwave field excites the propagating spin waves that drive the motion of the magnetic skyrmion from the source region to the drain region. The magnetic skyrmion at the source region is generated by applying a spin-polarized current with an additional microwave current. Our study shows that the motion of the magnetic skyrmion, which is governed by the gate voltage, can also be controlled by the frequency and amplitude of the microwave field exciting the propagating spin waves. Meanwhile, it shows that the microwave current with a certain frequency can significantly reduce the critical current density required for the nucleation of the magnetic skyrmion. The results indicate that the proposed microwave-field-driven and microwave-current-assisted methods are the effective approaches for building the skyrmion-based transistor-like device. The magnetic skyrmion-based transistor-like device operated and controlled by the microwave field and current will be beneficial for the future skyrmion-based spintronic circuits.

\begin{figure}[t]
\centerline{\includegraphics[width=0.5\textwidth]{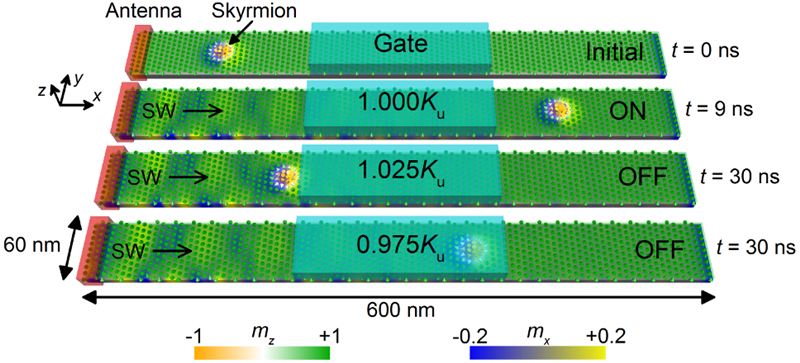}}
\caption{The microwave-driven magnetic skyrmion-based transistor-like device at different working states (initial state, ON state, and OFF state). The cones represent the magnetization, of which the out-of-plane component $m_z$ is denoted by the orange-white-green color scale and the in-plane component $m_x$ is denoted by the blue-gray-yellow color scale.}
\label{FIG1}
\end{figure}

\section{Model and simulation details}
\label{se:Methods}
The three-dimensional (3D) micromagnetic simulation is performed by using the 1.2a5 release of the Object Oriented MicroMagnetic Framework (OOMMF) software developed at the National Institute of Standards and Technology (NIST)~\cite{OOMMF}. The simulation is handled by the OOMMF extensible solver (OXS) objects of the standard OOMMF distribution with the OXS extension module for including the interface-induced Dzyaloshinskii-Moriya interaction (DMI), that is, the Oxs$\_$DMExchange6Ngbr class~\cite{Rohart_PRB2013,DMI_Dzyaloshinskii,DMI_Moriya}.

The interfacial DMI in the ultra-thin magnetic film with perpendicular magnetic anisotropy (PMA) placed on the heavy-metal substrate with high spin-orbit coupling is expressed as~\cite{Rohart_PRB2013}
\begin{equation}
E_{\rm{DM}}=\sum_{\langle i,j\rangle}d_{\rm{DM}}(\bi{u}_{ij}\times\hat{\bi{z}})\cdot(\bi{S}_{i}\times\bi{S}_{j}),
\label{DMIEnergy}
\end{equation}
with $d_{\rm{DM}}$ the DMI coupling energy,
$\bi{u}_{ij}$ the unit vector between spins $\bi{S}_{i}$ and $\bi{S}_{j}$,
and $\hat{\bi{z}}$ the normal to the interface, oriented from the heavy-metal substrate to the magnetic layer.
In the continuous micromagnetic model, the DMI reads
\begin{align}
E_{\rm{DM}}=b\int D[&(m_{x}\frac{\partial m_{z}}{\partial x}-m_{z}\frac{\partial m_{x}}{\partial x}) \notag \\
+(&m_{y}\frac{\partial m_{z}}{\partial y}-m_{z}\frac{\partial m_{y}}{\partial y})]d^{2}\bi{r},
\label{DMIEnergyMicro}
\end{align}
with $D$ the continuous DMI constant,
$b$ the magnetic sample thickness.
$m_{x}$, $m_{y}$ and $m_{z}$ are the components of the reduced magnetization $\bi{m}$, where $\bi{m}=\bi{M}/M_{\rm{S}}$.
The link between $D$ and $d_{\rm{DM}}$ is $d_{\rm{DM}}/ab$ for a ($001$) interface and $d_{\rm{DM}}\sqrt{3}/ab$ for a ($111$) interface with $a$ the atomic distance, of which the latter case is employed in this paper.

The 3D time-dependent magnetization dynamics in the simulation is controlled by the Landau-Lifshitz-Gilbert (LLG) equation including the spin-transfer torque (STT) term~\cite{OOMMF,LLGSTT}.
Specifically, when no spin-polarized current is injected into the simulated system, that is, the STT term is deactivated, the LLG equation reads
\begin{equation}
\frac{d\bi{M}}{dt}=-\gamma_{\rm{0}}\bi{M}\times\bi{H}_{\rm{eff}}+\frac{\alpha}{M_{\rm{S}}}(\bi{M}\times\frac{d\bi{M}}{dt}),
\label{LLG}
\end{equation}
where $\bi{M}$ is the magnetization,
$\bi{H}_{\rm{eff}}$ is the effective field,
$t$ is the time,
$\alpha$ is the Gilbert damping coefficient,
and $\gamma_{\rm{0}}$ is the gyromagnetic ratio.
The effective field $\bi{H}_{\rm{eff}}$ is expressed as follows
\begin{equation}
\bi{H}_{\rm{eff}}=-\mu_{0}^{-1}\frac{\delta \varepsilon}{\delta \bi{M}}.
\label{EffectiveField}
\end{equation}
The average energy density $\varepsilon$ is a function of $\bi{M}$ specified by
\begin{align}
\varepsilon=A[\nabla(\frac{\bi{M}}{M_{\rm{S}}})]^{2}&-K\frac{(\bi{n}\cdot\bi{M})^{2}}{M_{\rm{S}}^{2}}-\mu_{0}\bi{M}\cdot\bi{H}\notag \\
&-\frac{\mu_{0}}{2}\bi{M}\cdot\bi{H}_{\rm{d}}(\bi{M})+\varepsilon_{\rm{DM}},
\label{EnergyDensity}
\end{align}
where $A$ and $K$ are the exchange and anisotropy energy constants, respectively.
$\bi{H}$ and $\bi{H}_{\rm{d}}(\bi{M})$ are the applied and magneto-static self-interaction fields while $M_{\rm{S}}=|\bi{M}(r)|$ is the saturation magnetization.
$\varepsilon_{\rm{DM}}$ is the energy density of the DMI, which has the form
\begin{equation}
\varepsilon_{\rm{DM}}=\frac{D}{M_{\rm{S}}^{2}}(M_{z}\frac{\partial M_{x}}{\partial x}+M_{z}\frac{\partial M_{y}}{\partial y}-M_{x}\frac{\partial M_{z}}{\partial x}-M_{y}\frac{\partial M_{z}}{\partial y}),
\label{DMIDensity}
\end{equation}
where the $M_x$, $M_y$ and $M_z$ are the components of the magnetization $\bi{M}$.
The five terms at the right side of Eq.~(\ref{EnergyDensity}) correspond to the exchange energy, the anisotropy energy, the applied field (Zeeman) energy, the magneto-static (demagnetization) energy and the DMI energy, respectively.
For the simulated system, the spin-polarized current with the injection of current-perpendicular-to-the-plane (CPP) geometry is considered. The in-plane spin transfer torque is written as~\cite{Sampaio_NNANO2013}
\begin{equation}
\tau_{\rm{in-plane}}=-\frac{u}{b}\bi{m}\times(\bi{m}\times\bi{p}),
\label{CPPSTTtorqueIP}
\end{equation}
where $u=\frac{\gamma_{0}\hbar jP}{2\mu_{0}eM_{\rm{S}}}$,
$j$ is the current density,
$P$ is the spin polarization,
$\bi{p}=-\hat{z}$ is the unit electron polarization direction,
$b$ is the thickness of the ferromagnetic layer.
Thus, the LLG equation (Eq.~\ref{LLG}) of magnetization motion augmented with STT terms reads
\begin{equation}
\frac{d\bi{m}}{dt}=-\gamma_{0}\bi{m}\times\bi{h}_{\rm{eff}}+\alpha(\bi{m}\times\frac{d\bi{m}}{dt})-\frac{\gamma_{0}\hbar jP}{2\mu_{0}ebM_{\rm{S}}}[\bi{m}\times(\bi{m}\times\bi{p})],
\label{LLGSCPP}
\end{equation}
where $\bi{h}_{\rm{eff}}$ is the reduced effective field, that is, $\bi{h}_{\rm{eff}}=\bi{H}_{\rm{eff}}/M_{\rm{S}}$.
The models built in the micromagnetic simulation are divided into regular cells with the constant size of $2$ nm $\times$ $2$ nm $\times$ $1$ nm, which allows for a trade-off between numerical accuracy and computational efficiency.
The Oersted field is neglected in the simulation for simplicity due to its minor contribution to the effective field.

For micromagnetic simulations, the parameters of the magnetic layer are adopted from Refs.~\onlinecite{Fert_NNANO2013,Sampaio_NNANO2013,Tomasello_SREP2014}: Gilbert damping coefficient $\alpha = 0.02$; gyromagnetic ratio $\gamma = -2.211 \times 10^{5}$ m A$^{-1}$ s$^{-1}$; saturation magnetization $M_{\rm{S}} = 580$ kA m$^{-1}$; intralayer exchange stiffness $A = 15$ pJ m$^{-1}$; DMI constant $D = 0 \sim 4$ mJ m$^{-2}$; PMA $K_{\rm{u}} = 0.8$ MJ m$^{-3}$; and spin polarization rate $P=0.4$ unless otherwise specified.

The stable magnetic skyrmion stabilized by the interface-induced DMI is the hedgehog-like skyrmion, which has a radial in-plane magnetization profile.
It is characterized by the Pontryagin number $Q$~\cite{Yan_NCOMMS2014,Xichao_SREP2015B}, namely the topological charge in the planar system, which is defined by
\begin{equation}
Q=\int dxdy\rho_{\rm{sky}}(x),
\label{PontryaginNum}
\end{equation}
where the $\rho_{\rm{sky}}$ reads
\begin{equation}
\rho_{\rm{sky}}(x)=-\frac{1}{4\pi}\bi{m}(x)\cdot(\partial_{x}\bi{m}(x)\times\partial_{y}\bi{m}(x)).
\label{rho_sky}
\end{equation}
The number $Q$ is referred to as the skyrmion number.
When the background magnetization and the skyrmion core are pointing in the $+z$-direction and the $-z$-direction, respectively, the skyrmion number $Q$ equals $+1$.
Otherwise, when the background magnetization and the skyrmion core are pointing in the $-z$-direction and the $+z$-direction, respectively, the skyrmion number $Q$ equals $-1$.
In this paper, as we are assuming that the background magnetization is aligned along the $+z$-direction at the initial state, thus the skyrmion number of the relaxed (stable/metastable) skyrmion equals one, $Q=+1$.

\begin{figure}[t]
\centerline{\includegraphics[width=0.5\textwidth]{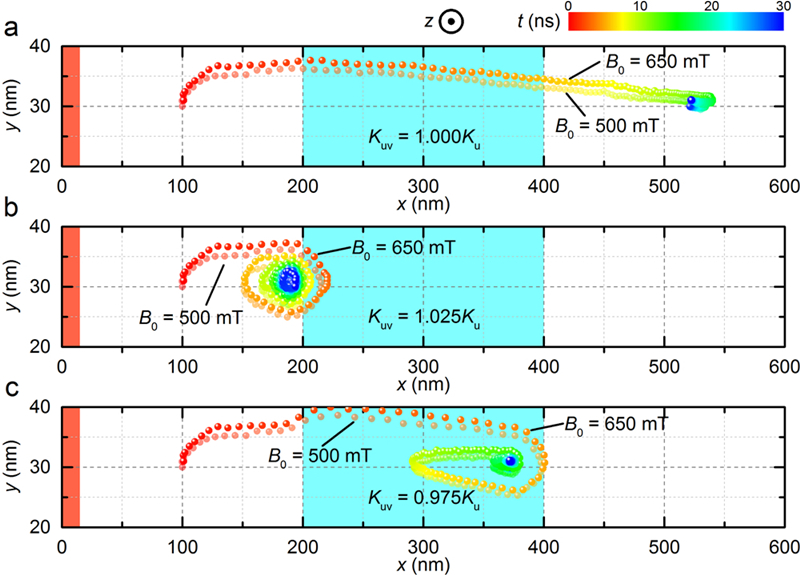}}
\caption{The trajectories of the magnetic skyrmion at the ON and OFF states driven by the microwave with different amplitudes.
(a) The magnetic skyrmion-based transistor-like device is in the ON state, where the microwave antenna is turned on and the voltage gate is turned off. $B_{\rm{0}} = 500$ mT or $650$ mT, $f = 75$ GHz, similar hereinafter. The PMA in the voltage-gated region ($K_{\rm{uv}}$) equals that of the outside region ($K_{\rm{u}}$).
(b) The magnetic skyrmion-based transistor-like device is in the OFF state, where the microwave antenna and the voltage gate are turned on.
The PMA in the voltage-gated region is larger than that of the outside region ($K_{\rm{uv}} = 1.025K_{\rm{u}}$).
(c) The magnetic skyrmion-based transistor-like device is in the OFF state, where the microwave antenna and the voltage gate are turned on.
The PMA in the voltage-gated region is smaller than that of the outside region ($K_{\rm{uv}} = 0.975K_{\rm{u}}$). The dots denote the center of the magnetic skyrmion. The simulation time is $30$ ns, which is indicated by the color scale. The red and blue regions stand for the regions where the microwave antenna and the voltage gate are deployed, respectively.
}
\label{FIG2}
\end{figure}

\begin{figure}[t]
\centerline{\includegraphics[width=0.5\textwidth]{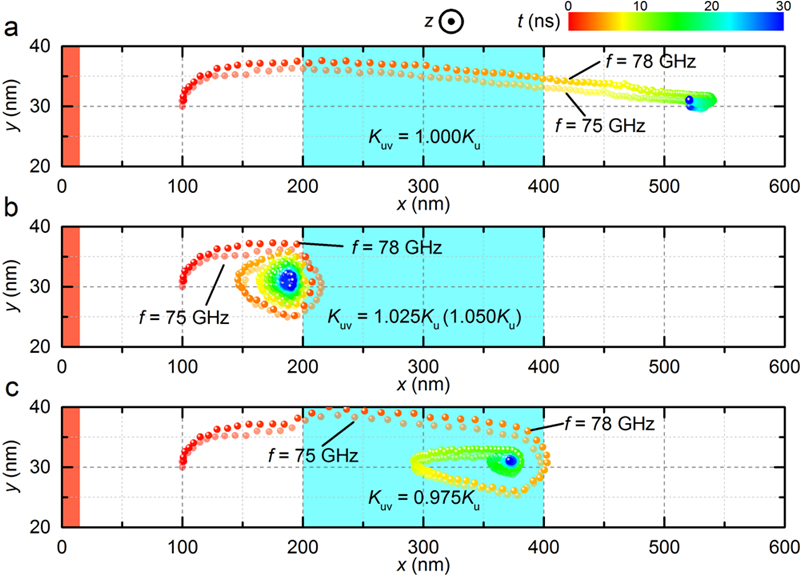}}
\caption{The trajectories of the magnetic skyrmion at the ON and OFF states driven by the microwave with different frequencies.
(a) The magnetic skyrmion-based transistor-like device is in the ON state, where the microwave antenna is turned on and the voltage gate is turned off. $B_{\rm{0}}=500$ mT, $f=75$ or $78$ GHz, similar hereinafter.
The PMA in the voltage-gated region ($K_{\rm{uv}}$) equals that of the outside region ($K_{\rm{u}}$).
(b) The magnetic skyrmion-based transistor-like device is in the OFF state, where the microwave antenna and the voltage gate are turned on.
The PMA in the voltage-gated region is larger than that of the outside region (when $f=75$ GHz, $K_{\rm{uv}}=1.025K_{\rm{u}}$, while when $f=78$ GHz, $K_{\rm{uv}}=1.050K_{\rm{u}}$).
(c) The magnetic skyrmion-based transistor-like device is in the OFF state, where the microwave antenna and the voltage gate are turned on.
The PMA in the voltage-gated region is smaller than that of the outside region ($K_{\rm{uv}}=0.975K_{\rm{u}}$). The dots denote the center of the magnetic skyrmion. The simulation time is $30$ ns, which is indicated by the color scale. The red and blue regions stand for the regions where the microwave antenna and the voltage gate are deployed, respectively.
}
\label{FIG3}
\end{figure}

\begin{figure}[t]
\centerline{\includegraphics[width=0.5\textwidth]{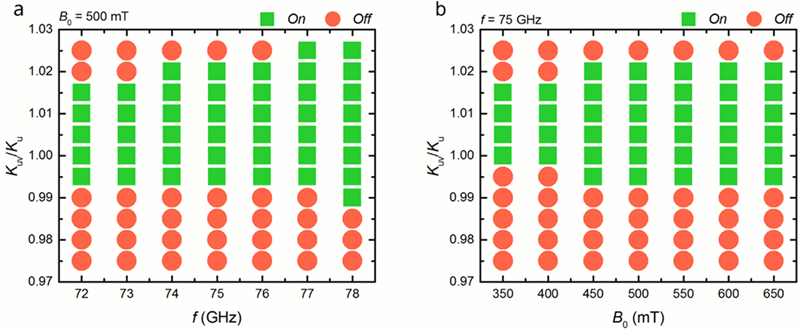}}
\caption{The working windows of the microwave-driven skyrmion-based transistor-like device.
(a) The working window at different gate voltages and microwave antenna excitation frequencies with a fixed microwave antenna excitation amplitude of $500$ mT.
(b) The working window at different gate voltages and microwave antenna excitation amplitudes with a fixed microwave antenna excitation frequency of $75$ GHz.
The square denotes the ON state, that is, the magnetic skyrmion passes the voltage-gated region moving from the source side to the drain side. The circle denotes the OFF state, that is, the magnetic skyrmion cannot pass the voltage-gated region and stops at the rest of the drain side.}
\label{FIG4}
\end{figure}

\section{Results and discussion}
\label{se:Results}

\subsection{Magnetic skyrmion-based transistor-like device driven by a microwave field}
\label{se:Microwave-Driving}
First, we study the magnetic skyrmion-based transistor-like device operated by a microwave field under the framework of micromagnetics (see Sec.~\ref{se:Methods} for the modeling details). As shown in Fig.~\ref{FIG1}, this device is basically constructed by a nanotrack with the size of $600$ nm $\times60$ nm $\times1$ nm, which has a PMA value of $K_{\rm{u}}$ $=$ $0.8$ MJ m$^{-3}$ and a Dzyaloshinskii-Moriya interaction (DMI) constant of $D=3.5$ mJ m$^{-2}$. The magnetic nanotrack is sandwiched between the voltage gate electrode and the heavy-metal substrate, where a voltage-gated region is between $x = 200$ nm and $x = 400$ nm. The PMA value within the voltage-gated region $K_{\rm{uv}}$ can be adjusted by applying an electric field $E_{\rm{gate}}$ based on the relationship of $K_{\rm{uv}}=K_{\rm{u}}+\Delta K_{\rm{uv}}E_{\rm{gate}}$~\cite{Maruyama_NNANO2009, Shiota_APE2011,Schellekens_NCOMMS2012}, where the transition regions between $K_{\rm{u}}$ and $K_{\rm{uv}}$ span 10 nm. The regions in the nanotrack at the left and right sides of the voltage-gated region are referred to as the source and drain sides, respectively. A sinusoidal microwave magnetic field $B_{\rm{0}}\sin(2\pi ft)$ along y-direction has been locally applied to generate spin waves at the end of the nanotrack ($0$ nm $< x <$ $15$ nm). $B_{\rm{0}}$ is the microwave amplitude and $f$ is the microwave frequency. A similar scheme has been used in Refs.~\onlinecite{Bance_JAP2008,Xing_IOP2015}. The magnetic nanotrack is almost magnetized along $+z$-direction in our setup. A magnetic skyrmion is created and relaxed at the source side of the nanotrack ($x = 100$ nm) by a skyrmion injector which can be fabricated by placing a magnetic tunnel junction (MTJ) upon the source side of the nanotrack~\cite{Koshibae_JJAP2015}.

Figure~\ref{FIG1} illustrates the initial, ON and OFF states of the magnetic skrymion-based transistor-like device. At the initial state, both the microwave antenna and the voltage gate are turned off, the magnetic skyrmion remains in its position on the source side at $x = 100$ nm. At the ON state, the antenna is turned on ($B_{\rm{0}}=500$ mT, $f = 75$ GHz) but the voltage gate is turned off. The microwave field pulse applied at the left end of the nanotrack excites spin waves propagating toward the drain side of the nanotrack, driving the magnetic skyrmion into motion. The moving skyrmion passes the voltage-gated region and reaches the drain side of the nanotrack at $t = 9$~ns, which can be detected by the skyrmion reader at the drain side~\cite{Koshibae_JJAP2015}. At the OFF state, both the microwave antenna and the voltage gate are turned on. The spin waves excited by the microwave field drive the magnetic skyrmion moving toward the right end of the nanotrack. The gate voltage results in the change of the PMA value $K_{\rm{uv}}$ in the voltage-gated region, leading to the stop of skyrmion when it approaches the gate-induced potential barrier. As shown in Fig.~\ref{FIG1}, when the PMA value in the voltage-gated region is larger than that of the intrinsic value, i.e. $K_{\rm{uv}} = 1.025K_{\rm{u}}$, the skyrmion stops at the potential barrier at the boundary between the source side and the voltage-gated region. When the PMA value in the voltage-gated region is smaller than that of the intrinsic value, i.e. $K_{\rm{uv}} = 0.975K_{\rm{u}}$, the skyrmion stops at the potential barrier at the boundary between the voltage-gated region and the drain side. For both OFF states, the skyrmion reaches the equilibrium state within $t = 30$ ns under the driving force from the microwave-induced spin waves and the repulsive force from the potential barrier.

\begin{figure}[t]
\centerline{\includegraphics[width=0.5\textwidth]{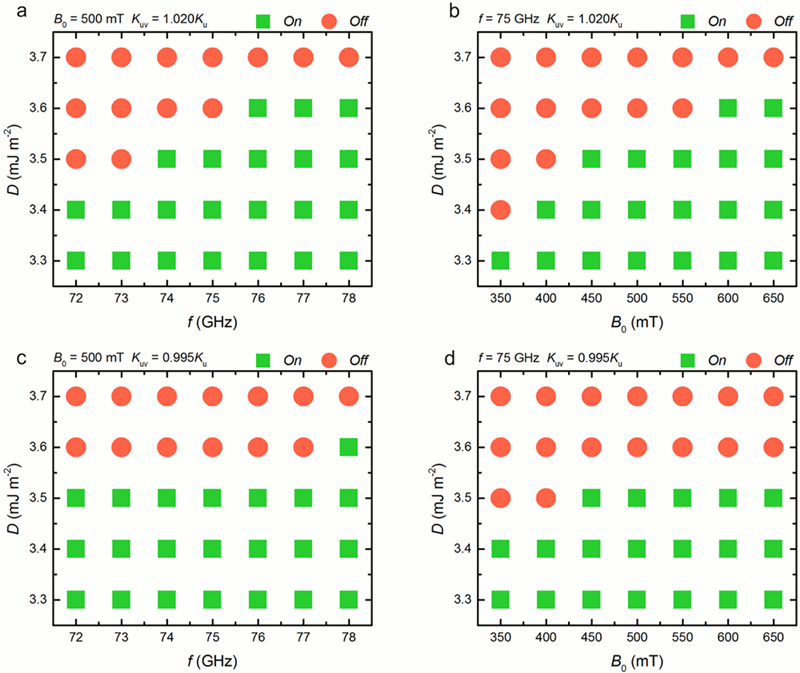}}
\caption{The effect of Dzyaloshinskii-Moriya interaction (DMI) on the working window of the microwave-driven skyrmion-based transistor-like device.
(a) The working window at different DMI and microwave antenna excitation frequencies with fixed $K_{\rm{uv}}=1.020K_{\rm{u}}$ and $B_{\rm{0}}=500$ mT.
(b) The working window at different DMI and microwave antenna excitation amplitudes with fixed $K_{\rm{uv}}=1.020K_{\rm{u}}$ and $f=75$ GHz.
(c) The working window at different DMI and microwave antenna excitation frequencies with fixed $K_{\rm{uv}}=0.995K_{\rm{u}}$ and $B_{\rm{0}}=500$ mT.
(d) The working window at different DMI and microwave antenna excitation amplitudes with fixed $K_{\rm{uv}}=0.995K_{\rm{u}}$ and $f=75$ GHz.
The square denotes the ON state, that is, the magnetic skyrmion passes the voltage-gated region moving from the source side to the drain side. The circle denotes the OFF state, that is, the magnetic skyrmion cannot pass the voltage-gated region and stops at the rest of the drain side.}
\label{FIG5}
\end{figure}

\begin{figure}[t]
\centerline{\includegraphics[width=0.5\textwidth]{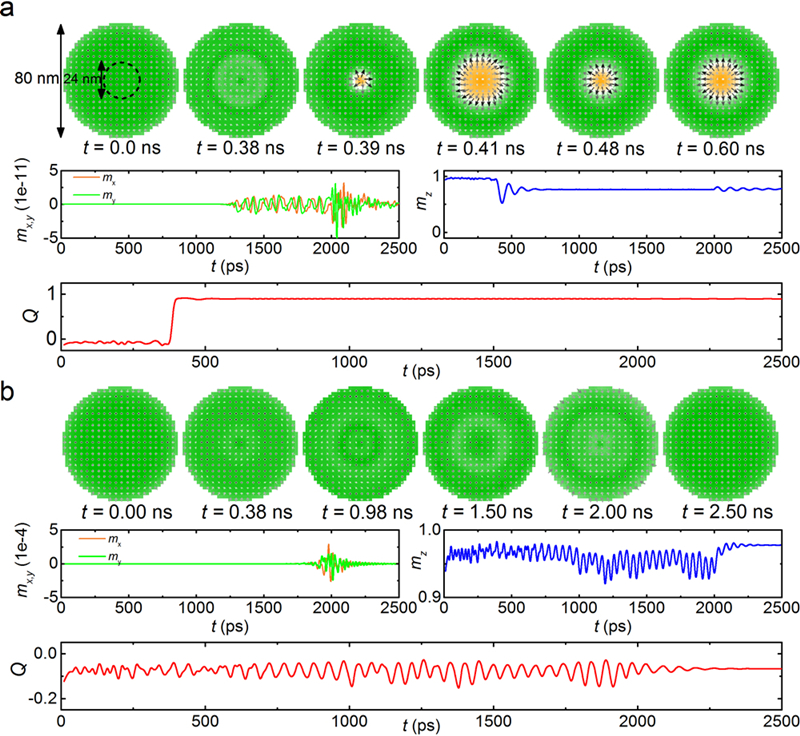}}
\caption{The microwave-current-assisted nucleation of the magnetic skyrmion.
(a) Selected time evolution of the magnetization distribution, time-dependent $m_x, m_y, m_z$, and skyrmion number $Q$ for the case of the skyrmion nucleation assisted by a microwave current ($j_{\rm{0}}=1\times10^{12}$ A m$^{-2}$) with a selected frequency of $f_{\text{AC}}=53$ GHz. The diameter of the nanodisk equals $80$ nm. The diameter of the current injection region equals $24$ nm, which is denoted by the dashed circle in
(a). A magnetic skyrmion can be nucleated successfully even $j_{\rm{DC}}$ ($1.8\times10^{12}$ A m$^{-2}$) is lower than the threshold current density, $j_{\rm{th}}=2\times10^{12}$ A m$^{-2}$.
(b) Selected time evolution of the magnetization distribution, time-dependent $m_x, m_y, m_z$, and skyrmion number $Q$ for the case of the skyrmion nucleation by applying only a DC spin current of $j_{\rm{DC}}=1.8\times10^{12}$ A m$^{-2}$.
In the absence of the microwave current, the DC spin current itself of $j_{\rm{DC}}=1.8\times10^{12}$ A m$^{-2}$ is unable to create the magnetic skyrmion. The initial magnetization state of the nanodisk is along the $+z$-direction, and the polarization of the spin current is aligned along the $-z$-direction.}
\label{FIG6}
\end{figure}

\begin{figure}[t]
\centerline{\includegraphics[width=0.5\textwidth]{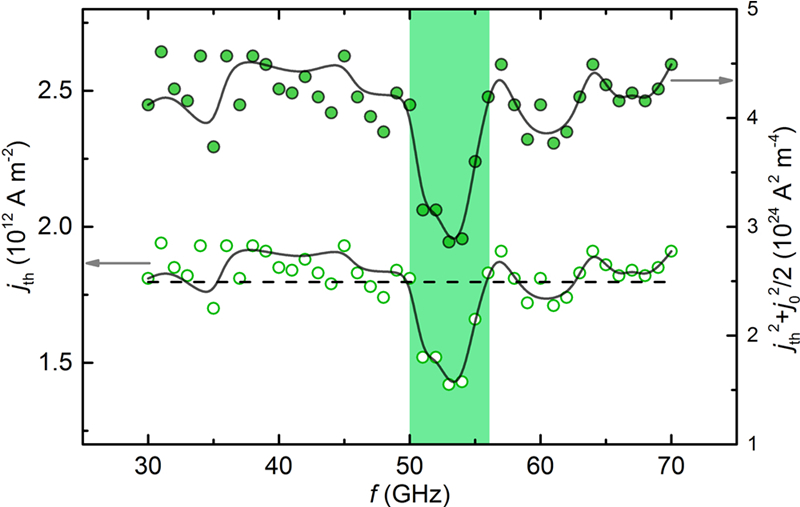}}
\caption{The effect of the frequency of the microwave current on the threshold current density required to create the magnetic skyrmion. The solid and open circles represent $j_{\rm{th}}^2+j_{\rm{0}}^2/2$ and the threshold current density $j_{\rm{th}}$ respectively. The amplitude $j_{\rm{0}}$ is fixed to be $1.3\times10^{12}$ A m$^{-2}$. The step of the frequency of microwave current is $1$ GHz. The lines are guides to the eye. The assistance effect of the microwave current varies with its frequency $f_{\text{AC}}$, and the most effective frequency is indicated by green shadow. The dash line indicates the average value of the $j_{\rm{th}}$.}
\label{FIG7}
\end{figure}

\begin{figure}[t]
\centerline{\includegraphics[width=0.5\textwidth]{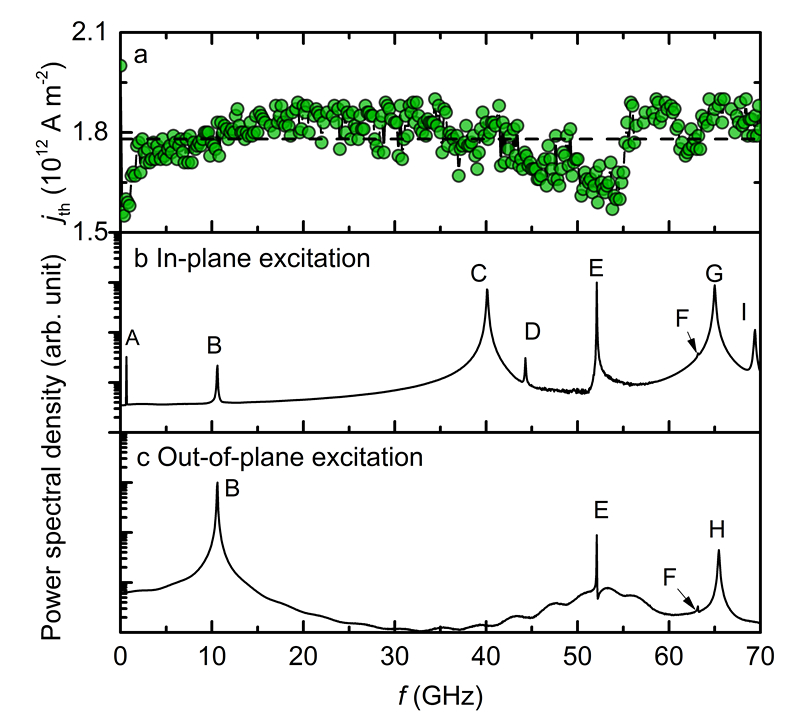}}
\caption{The threshold current density $j_{\rm{th}}$ as a function of the microwave current frequency $f_{\text{AC}}$ with a fixed amplitude of $j_{\rm{0}}=1.0\times10^{12}$ A m$^{-2}$ and normalized power spectral densities.
(a) The threshold current density $j_{\rm{th}}$ as a function of the microwave current frequency $f_{\text{AC}}$. The dash line indicates the average value of the $j_{\rm{th}}$.
(b) Power spectral densities of SK for the in-plane excitation.
(c) Power spectral densities of SK for the out-of-plane excitation.}
\label{FIG8}
\end{figure}

Figure~\ref{FIG2} shows the trajectories of the skyrmion motion in the magnetic skrymion-based transistor-like device driven by the microwave-induced spin waves at the ON and OFF states  with different amplitudes ($B_{\rm{0}}=500$ mT or $650$ mT, $f=75$ GHz). As shown in Fig.~\ref{FIG2}a, the magnetic skyrmion-based transistor-like device is in the ON state, where $K_{\rm{uv}}=K_{\rm{u}}$. The skyrmion moves from the source side to the drain side within $10$ ns. It is worth mentioning that the transverse motion of the skyrmion in the $y$-direction is caused by the skyrmion Hall effect~\cite{Xichao_NCOMMS2015}. Due to the decay of the microwave-induced spin waves and the repulsion from the edge at the right end of the nanotrack, the magnetic skyrmion is finally relaxed at the drain side at $t=30$ ns. For $K_{\rm{uv}}=1.025K_{\rm{u}}$, the magnetic skyrmion-based transistor-like device is in the OFF state, as shown in Fig.~\ref{FIG2}b. However, the magnetic skyrmion cannot surmount the gate-induced potential barrier. It can be seen that the center of the magnetic skyrmion can enter the voltage-gated region for a short time driven by the strong excited spin wave emitted from the microwave antenna. Nevertheless, the magnetic skyrmion is repelled by the potential barrier. The combined effects of the driving force supplied from the microwave antenna, the repulsive force provided by the voltage gate and the Magnus force exerted on the magnetic skyrmion lead to the motion of the magnetic skyrmion in a spiral trajectory. The skyrmion eventually reaches equilibrium and stops near the source side boundary of voltage-gated region. Figure~\ref{FIG2}c shows the other OFF state, where $K_{\rm{uv}}=0.975K_{\rm{u}}$. The magnetic skyrmion driven by the microwave-induced spin waves moves toward the drain side. In contrast to the OFF state shown in Fig.~\ref{FIG2}b, the magnetic skyrmion passes the source side boundary of voltage-gated region, where a potential well is induced. However, the magnetic skyrmion cannot penetrate the potential well. Similar to the OFF state shown in Fig.~\ref{FIG2}b, the magnetic skyrmion moves in a spiral path, and finally stops near the boundary between the voltage-gated region and the drain side. In three cases, the frequency of the microwave is fixed at $f=75$ GHz, while two amplitudes of $B_{\rm{0}} = 500$ mT and $B_{\rm{0}} = 650$ mT are applied, respectively. It can be seen that the transverse shift of the skyrmion in the $y$-direction increases with the amplitude of the microwave. Indeed, the speed of the skyrmion increases with the amplitude of the microwave.

Figure~\ref{FIG3} shows the trajectories of the skyrmion in the magnetic skyrmion-based transistor-like device at the ON and OFF states driven by the microwave-induced spin waves with different frequencies ($B_{\rm{0}}=500$ mT, $f=75$ GHz or $78$ GHz). In three cases, the amplitude of the microwave is fixed at $B_{\rm{0}} = 500$ mT, while two frequencies of $f = 75$ GHz and $f = 78$ GHz are applied, respectively. It should be noted that in Fig.~\ref{FIG3}c, we set $K_{\rm{uv}} = 1.025K_{\rm{u}}$ for the case of $f = 75$ GHz, while $K_{\rm{uv}} = 1.050K_{\rm{u}}$ for the case of $f = 78$ GHz, in order to ensure the OFF state. It can be seen that the behaviors of the magnetic skyrmions are similar to these shown in Fig.~\ref{FIG2}. The transverse shift and the speed of the skyrmion increases with the frequency of the microwave.

Figure~\ref{FIG4} shows the working windows of the magnetic skyrmion-based transistor-like device driven and controlled by the microwave antenna and the voltage gate. As shown in Fig.~\ref{FIG4}a, the excitation field amplitude of the microwave antenna is fixed at $B_{\rm{0}}=500$ mT, while the excitation field frequency of the microwave antenna is varied in the range between $f=72$ GHz and $f=78$ GHz. Obviously, when the voltage gate is turned off, i.e. $K_{\rm{uv}}=K_{\rm{u}}$, the magnetic skyrmion moves from the source side to the drain side in a certain time and this transistor-like device is always in the ON state. When the voltage gate is turned on, which adjusts the $K_{\rm{uv}}$ to be in the range of $0.995K_{\rm{u}}$ and $1.015K_{\rm{u}}$, the magnetic skyrmion-based transistor-like device is still in the ON state, as the gate-induced potential barrier is not strong enough to stop the magnetic skyrmion from passing through. When the $K_{\rm{uv}}$ is further increased to be larger than $1.02K_{\rm{u}}$ or decreased to be smaller than $0.99K_{\rm{u}}$, the working state switches to the OFF state at $f=72$ GHz. However, it can be seen that when $f$ increases to $78$ GHz, the magnetic skyrmion under the stronger driving force provided by the microwave antenna can overcome the potential barrier induced by the voltage gate in the range of $K_{\rm{uv}}=0.99K_{\rm{u}}$ and $K_{\rm{uv}}=1.03K_{\rm{u}}$. Because the source side boundary of the voltage-gated region is closer to the antenna than the drain side boundary of the voltage-gated region, the skyrmion experiences stronger driving force when it is approaching the former boundary. Thus, the potential barrier at the boundary between the source side with lower PMA and the voltage-gated region with higher PMA is easier to be penetrated by the magnetic skyrmion at the same $B_{\rm{0}}$ and $f$. Similar results are shown in Fig.~\ref{FIG4}b, where the excitation field frequency of the microwave antenna is fixed at $f = 75$ GHz, while the excitation field amplitude of the microwave antenna is varied in the range between $B_{\rm{0}} = 350$ mT and $B_{\rm{0}} = 650$ mT.

Figure~\ref{FIG5} shows the working windows of the magnetic skyrmion-based transistor-like device at different DMI constants and microwave antenna parameters with a fixed voltage gate-induced PMA value $K_{\rm{uv}}$. Figs.~\ref{FIG5}a and \ref{FIG5}b show the working state at $K_{\rm{uv}} = 1.020K_{\rm{u}}$ as functions of the DMI constant $D$ and antenna frequency $f$, and as functions of DMI constant $D$ and antenna amplitude $B_{\rm{0}}$, respectively. Figure~\ref{FIG5}c and \ref{FIG5}d show the working state at $K_{\rm{uv}} = 0.995K_{\rm{u}}$ as functions of the DMI constant $D$ and antenna frequency $f$, and as functions of DMI constant $D$ and antenna amplitude $B_{\rm{0}}$, respectively. It can be seen that the magnetic skyrmion in a nanotrack with a smaller DMI constant $D$ is easier to overcome the potential barrier at given $K_{\rm{uv}}$, $B_{\rm{0}}$ and $f$.

\emph{The voltage gate-induced PMA gives rise to a change in the potential barrier for skyrmions, which results in the transistor-like action. At the same time, it is expected to result in the reflection of spin waves, which may modify the net force on skyrmions, thus also affecting the observed transistor-like action. The increasing the frequency increases the force on skyrmions, making it easier for skyrmions to cross the barrier. The spin wave with large frequency has large energy, resulting in the large velocity of the magnetic skyrmion. Then, for the same barrier and skyrmion, the spin wave with high frequency drives the magnetic skyrmion to pass the barrier easily. The increasing DMI also makes the magnetic skyrmion easier to overcome the barrier. This is because that the large DMI results in a large-size magnetic skyrmion and the number of magnetization interacting with the spin wave is increasing. Then, the driving force is enhanced.}


\subsection{Nucleation of the magnetic skyrmion assisted by a microwave current}
\label{se:Microwave-Nucleation}
With respect to the generation of the magnetic skyrmion at the source terminal of the magnetic skyrmion-based transistor-like device, we also implement the microwave-current-assisted nucleation of the magnetic skyrmion. For the sake of low computational complexity, here we consider the model of a nanodisk instead of the nanotrack studied in last section. As shown in Fig.~\ref{FIG6}a, a magnetic nanodisk initially magnetized along $+z$-direction is built in the simulation, which has a diameter of $80$ nm and a thickness of $1$ nm. The current injection is injected through a nano-contact with a diameter $24$ nm in the central region of the nanodisk.

In order to create the magnetic skyrmion in the nanodisk, we utilize a microwave-assisted vertical spin-polarized current, which can be expressed as $j_{\rm{total}} = j_{\rm{0}}\sin{2\pi f_{\text{AC}}t}+j_{\rm{DC}}$. Here, $j_{\rm{0}}$ is the amplitude of the microwave current, $f_{\text{AC}}$ is the frequency of the microwave current, and $j_{\rm{DC}}$ is the DC current density. We focus on the threshold current density $j_{\rm{th}}$ which is defined as the minimal $j_{\rm{DC}}$ required to nucleate the magnetic skyrmion within a $2$ ns injection of spin current and microwave current. The applied current duration is fixed to $2$ ns in this work.

Figure~\ref{FIG6}a shows the nucleation process of the magnetic skyrmion in the presence of the microwave current. The DC current density $j_{\rm{DC}}$ is set as $j_{\rm{DC}} = 1.8\times10^{12}$ A m$^{-2}$, and the frequency of the microwave current is set as $f_{\text{AC}}=53$ GHz. The nucleation current is injected at $t=0$ ns. The skyrmion number $Q$ starts to fluctuate while the magnetization stays quasi-uniform with a spin wave transferring from the center to the edge, indicating the injection of energy. At $t=0.39$ ns, the magnetization of the central region reverses in a very short time ($\sim 0.01$ ns), and generates a stable magnetic skyrmion, resulting in a rapid jump of the skyrmion number from $Q\sim -0.1$ to $Q\sim 1$. Then, the skyrmion number stabilizes at $Q\sim 1$ and a skyrmion is created successfully. It should be noted that we also find the breathing of the magnetic skyrmion after its nucleation, which can be seen from the snapshots and the fluctuation of $m_z$ during the period of $t=0.39\sim 0.60$ ns. At $t=0.60$ ns, the magnetic skyrmion stops breathing and becomes a stable magnetic skyrmion. Figure~\ref{FIG6}b shows the case in the absence of the assistance of the microwave current. With the identical DC current density, no magnetic skyrmion is formed in the nanodisk in the absence of the microwave current. The skyrmion number $Q$ fluctuates slightly during the time within which the DC current is applied. The skyrmion number is equal to $Q\sim 0$ at $t=2.5$ ns. It indicates that the DC current density of $1.8\times10^{12}$ A m$^{-2}$ is smaller than the threshold current.

\begin{figure}[t]
\centerline{\includegraphics[width=0.5\textwidth]{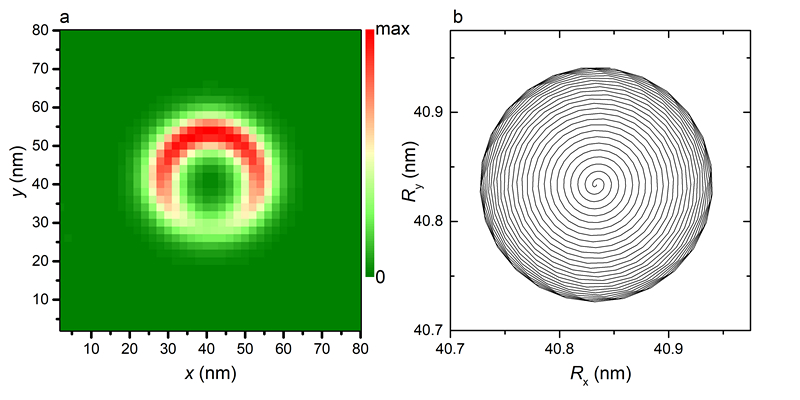}}
\caption{\emph{(a) Spatial distribution of PSD for peak A ($f=0.65$~GHz). (b) The trajectory of the guiding center of the magnetic skyrmion during the oscillation when in-plane excitation $B_\rm{0}\sin(2\pi f)$ with $B_\rm{0}=5$ mT and $f=0.65$~GHz is applied. For the method to calculate the guiding center of the magnetic skyrmion, we refer to Ref.~\onlinecite{Li2017}.}}
\label{FIG9}
\end{figure}
\begin{figure}[t]
\centerline{\includegraphics[width=0.5\textwidth]{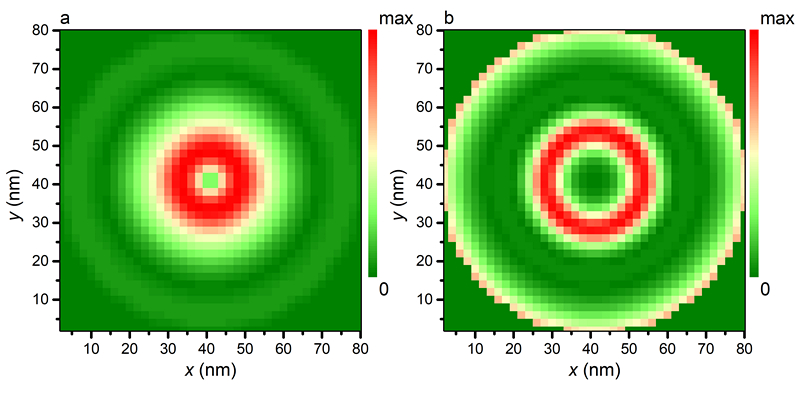}}
\caption{\emph{Spatial distributions of PSD for (a) peak E ($f=52.1$~GHz) and (b) peak G ($f=65$~GHz).}}
\label{FIG10}
\end{figure}

Figure~\ref{FIG7} shows the threshold current density $j_{\rm{th}}$ as functions of frequency $f_{\text{AC}}$ of the microwave current with a fixed microwave amplitude $j_{\rm{0}}=1.3\times10^{12}$ A m$^{-2}$. Since the power is in directly proportional to $j_{\rm{total}}^2$, we can obtain that the power is in direct proportional to $j_{\rm{th}}^2+j_{\rm{0}}^2/2$. Both the threshold current density and the corresponding $j_{\rm{th}}^2+j_{\rm{0}}^2/2$ have been shown in Fig.~\ref{FIG7}. As can be seen, the average $j_{\rm{th}}$ equals $1.8\times10^{12}$ A m$^{-2}$ approximately when the microwave current is injected. There is an obvious frequency range, $50$ GHz $\sim 56$ GHz, where the required current density is much less than the average one. As a result, the energy consumption in whole process is also reduced as the injection time is fixed to be $2$ ns. It means that the energy consumption can be reduced with modulating the frequency of the additional microwave current.

Figure~\ref{FIG8} shows normalized power spectral densities (PSDs) and the threshold current density $j_{\rm{th}}$ as a function of the microwave current frequency $f_{\text{AC}}$ with a fixed amplitude of $j_{\rm{0}} = 1.0\times10^{12}$ A m$^{-2}$. And the step size of the microwave current frequency is adopted as $0.2$ GHz. In order to obtain the PSDs of skyrmion, the equilibrium magnetization configuration $\bi{m}(0)$ has been calculated firstly. Then a sinc-function field $B_0 \sin (2\pi f_B t)/(2\pi f_B t)$ with $B_0=0.5$ mT and $f_B=200$ GHz is applied along $x$-axis (in-plane). We record the time-dependent magnetization configuration $\bi{m}(t)$ every $\Delta t=2$ ps and analyze $\Delta m_x(t)$ with spatially resolved (SR) methods~\cite{Beg2016,Baker2016}, as shown in Fig.~\ref{FIG8}b. The SR method requires computation of discrete Fourier transforms at all spatial sampling points. The SR PSD is obtained by averaging the local PSDs. The PSDs in Fig.~\ref{FIG8}c are obtained by analyzing $\Delta m_z(t)$ when a sinc-function field is applied along $z$-axis (out-of-plane). 9 peaks (A-I) are identified from the PSDs for skyrmion. For the peaks A, D, and E, the visible decreases of threshold current density are found. To identify the modes contributing to the decreases of critical current, we excite the oscillation of skyrmion  with a magnetic field $B_\rm{0}\sin(2\pi f)$ with $B_\rm{0}=5$ mT.
\emph{For peak A ($f=0.65$ GHz), the spatial distribution of PSD are shown in Fig.~\ref{FIG9}a. And a gyration of the guiding center of the magnetic skyrmion is observed, as shown in Fig.~\ref{FIG9}b. The mode for peak A is gyrotropic mode, resulting the decrease of the critical current density. For peak B ($f=10.6$ GHz), the mode is breathing mode. The other modes are standing spin wave modes. For these standing spin wave modes, resonance E leads to the visible reduce of the critical current density. We compare the spatial distributions of PSD for peaks E and peak G, as shown in Fig.~\ref{FIG10}. It shows that the oscillation for peak E is focused in the skyrmion center region while the oscillation for peak G is not focused in the skyrmion center region and the magnetization in the edge also oscillates obviously. Then, there is a visible decrease of the critical current density for peak E while no reduction for the case of peak G. For the microwave with low frequency, when the frequency approaches the eigenfrequency of the gyrotropic mode, the critical current density decreases. A recent work \cite{Li2017} shows that the stable skyrmion lattice can be created with the resonance excitation when the field frequency is equal to the eigenfrequency of the gyrotropic mode. For the microwave with high frequency, the critical current decreases when the frequency approaches the resonance frequency which causes the oscillation focused in the skyrmion center region.}

\section{Conclusions}
\label{se:Conclusions}
We have proposed a transistor-like function of magnetic skyrmion operated and controlled by microwaves. It should be mentioned that the magnetic skyrmion-based transistor-like device in this work has no amplification function so far. It is demonstrated that the microwave field can lead to the motion of the magnetic skyrmion by exciting propagating spin waves, where the motion of the magnetic skyrmion is governed by a gate voltage. The creation of the magnetic skyrmion at the source region is also assisted by a microwave current. It shows that the microwave current can reduce the threshold current density required for the nucleation of the magnetic skyrmion with a fixed time, and the reduction is significant when its frequency is close to the frequency of gyration mode. The critical nucleation current decreases from $2.0\times10^{12}$ A m$^{-2}$ to $1.58\times10^{12}$ A m$^{-2}$ if there is a microwave current with a few GHz assists the nucleation. And $25\%$ energy is saved. Although the energy efficiency of the generation of the DC and AC current is not taken into considerations for the micromagnetic simulation, the remarkable energy saved in the nucleation process indicates that the microwave-assisted method is promising. Our results on the magnetic skyrmion-based transistor-like device operated and controlled by the microwave field and microwave current might be useful in the design of future skyrmion-based spintronic circuits.

\begin{acknowledgements}
\emph{X.Z. was supported by the JSPS RONPAKU (Dissertation Ph.D.) Program. W.S.Z. acknowledges the support by the projects from the Chinese Postdoctoral Science Foundation (No. 2015M570024), National Natural Science Foundation of China (Projects No. 61501013, No. 61471015 and No. 61571023), Beijing Municipal Commission of Science and Technology (Grant No. D15110300320000), and the International Collaboration Project (No. 2015DFE12880) from the Ministry of Science and Technology of China. Y.Z. acknowledges the support by the President's Fund of CUHKSZ, the National Natural Science Foundation of China (Grant No. 11574137), and Shenzhen Fundamental Research Fund (Grant Nos. JCYJ20160331164412545 and JCYJ20170410171958839).}
\end{acknowledgements}


%
%
%
%
%
%

\end{document}